\documentclass[aps,floatfix,twocolumn,superscriptaddress]{revtex4-1}

\usepackage[english]{babel}
\usepackage{amsmath}
\usepackage{graphicx}
\usepackage{float}
\usepackage{bm}
\usepackage{multirow}
\usepackage{dcolumn}
\usepackage{color}
\usepackage{hyperref}

\newcommand{\icm}{\ensuremath{~\textrm{cm}^{-1}}}

\begin{document}

\title{Infrared study of the interplay of charge, spin, and lattice excitations in the magnetic topological insulator EuIn$_2$As$_{2}$}

\author{Bing Xu}
\email[]{bing.xu@unifr.ch}
\affiliation{University of Fribourg, Department of Physics and Fribourg Center for Nanomaterials, Chemin du Mus\'{e}e 3, CH-1700 Fribourg, Switzerland}

\author{P. Marsik}
\author{S. Sarkar}
\author{F. Lyzwa}
\affiliation{University of Fribourg, Department of Physics and Fribourg Center for Nanomaterials, Chemin du Mus\'{e}e 3, CH-1700 Fribourg, Switzerland}

\author{Y. Zhang}
\author{B. Shen}
\email[]{shenbing@mail.sysu.edu.cn}
\affiliation{Sate Key Laboratory of Optoelectronic Materials and Technologies, School of Physics, Sun Yat-Sen University, Guangzhou, Guangdong 510275, China}

\author{C. Bernhard}
\email[]{christian.bernhard@unifr.ch}
\affiliation{University of Fribourg, Department of Physics and Fribourg Center for Nanomaterials, Chemin du Mus\'{e}e 3, CH-1700 Fribourg, Switzerland}

\date{\today}

%
%
\begin{abstract}
We report an infrared spectroscopy study of the axion topological insulator candidate EuIn$_2$As$_2$ for which the Eu moments exhibit an A-type antiferromagnetic (AFM) order below $T_N \simeq 18~\mathrm{K}$. The low energy response is composed of a weak Drude peak at the origin, a pronounced infrared-active phonon mode at 185~cm$^{-1}$ and a free carrier plasma edge around 600~cm$^{-1}$. The interband transitions start above 800~cm$^{-1}$ and give rise to a series of weak absorption bands at 5\,000 and 12\,000~cm$^{-1}$ and strong ones at 20\,000, 27\,500 and 32\,000~cm$^{-1}$. The AFM transition gives rise to pronounced anomalies of the charge response in terms of a cusp-like maximum of the free carrier scattering rate around $T_N$ and large magnetic splittings of the interband transitions at 5\,000 and 12\,000~cm$^{-1}$. The phonon mode at 185~cm$^{-1}$ has also an anomalous temperature dependence around $T_N$ which suggests that it couples to the fluctuations of the Eu spins. The combined data provide evidence for a strong interaction amongst the charge, spin and lattice degrees of freedom.
\end{abstract}


\maketitle

%
%
\section{Introduction}
Topological insulators represent a new quantum state of matter for which gapless conducting states can develop at the surface of materials that are insulating in the bulk~\cite{Hasan2010RMP,Qi2011RMP,Haldane2017RMP}. These surface states disperse linearly around so-called Dirac points where they are protected by time-reversal symmetry and thus robust against perturbations (scattering). In the presence of an additional magnetic order, which breaks this time-reversal symmetry, a gap is opened at this Dirac point~\cite{Tokura2019NRP}. In such magnetic topological materials, the combination of non-trivial band topology and magnetic order may give rise to the emergence of a variety of novel quantum phenomena~\cite{Wan2011PRB,Yu2010SC,Chang2013SC,Chang2015NM,Qi2008PRB,Essin2009PRL,Mong2010PRB,Xiao2018PRL,Varnava2018PRB,He2017SC,Belopolski2019SC}. In recent years, an increasing number of magnetic materials, which have an intrinsic magnetic order and topological electronic states in the stoichiometric compositions, has been theoretically predicted as magnetic topological insulators~\cite{Chen2014PRL,Otrokov2019PRL,Zhang2019PRL,Li2019SA}, magnetic Dirac semimetals~\cite{Tang2016NP,Hua2018PRB}, and magnetic Weyl semimetals~\cite{Wang2016PRL,Xu2018PRB}. Such materials would not only provide a clean platform to realize the exotic topological phenomena under time-reversal symmetry breaking, but also show great potential for applications in quantum technology.

Recently, a strong experimental focus has been on MnBi$_2$Te$_4$~\cite{Otrokov2019Nat,Cui2019PRB,Gong2019CPL,Yan2019PRM,Yan2019PRB,Deng2020SC,Liu2020NM,Ge2020NSR,Hu2020NC,Lee2019PRR,Zeugner2019CM,Vidal2019PRB,Chen2019PRX,Swatek2020PRB,Hao2019PRX,Li2019PRX,Nevola2020,Xu2021PRB}, which has been predicted to be an intrinsic antiferromagnetic (AFM) topological insulator for which different topological states can be realized in bulk crystals as well as in thin films~\cite{Otrokov2019PRL,Li2019SA,Zhang2019PRL}. EuIn$_2$As$_{2}$ is another promising candidate for an intrinsic magnetic topological material~\cite{Xu2019PRL}. Different from the layered van der Waals-type MnBi$_2$Te$_4$, EuIn$_2$As$_{2}$ has a three-dimensional structure and crystallizes in the hexagonal $P6_3/mmc$ (No. 194) space group, with alternating stacking of Eu$^{2+}$ and [In$_2$As$_2$]$^{2-}$ layers along the $c$-axis~\cite{Xu2019PRL}. The Eu spins exhibit an A-type antiferromagnetic order below $T_N \simeq$ 18~K where they are parallel oriented within each layer and an antiparallel between neighbouring layers (along the $c$-axis)~\cite{Goforth2008IC,Zhang2020PRB}. It has been predicted that the AFM order in EuIn$_2$As$_{2}$ gives rise to an axion insulator with non-trivial topological states that are strongly influenced by the orientation of the magnetic moments. A topological crystalline insulator phase with gapless surface states on the (100), (010), and (001) surfaces is expected for in-plane oriented magnetic moments, whereas a higher-order topological insulator phase with chiral hinge states is predicted if the magnetic moments are out-of-plane oriented~\cite{Xu2019PRL}. From an experimental perspective, angle resolved photoemission spectroscopy (ARPES) studies have confirmed that EuIn$_2$As$_{2}$ has hole-type Fermi pockets around the bulk Brillouin zone center~\cite{Regmi2020PRB,Zhang2020PRB,Sato2020PRR}, together with a heavily hole-doped surface state and an inversion of the bulk band in the AFM state~\cite{Sato2020PRR} that is consistent with the theoretical prediction~\cite{Xu2019PRL}. A negative magneto-resistance seen in magneto-transport measurements has provided evidence for a rather strong spin scattering of the carriers by the localized magnetic moments~\cite{Goforth2008IC,Zhang2020PRB} that may affect the above described topological states. Electron spin resonance measurements have revealed that the spin dynamics in the vicinity of $T_N$ is governed by short-range AFM correlations of the Eu spins~\cite{Rosa2012PRB}. An optical spectroscopy study, which can directly probe the dynamics of the charge carriers and provide information about their interplay with the spin and lattice degrees of freedom, has not been reported to date (to our best knowledge).

Here, we present an infrared spectroscopy study of EuIn$_2$As$_2$ which reveals a strong interaction of the charge carriers with the Eu spins in terms of a cusp-like maximum of the free carrier scattering rate around $T_N$ and a sizeable exchange splitting of the bulk valence bands below $T_N$. Moreover, we observe corresponding anomalies of an infrared-active phonon mode around 185\icm\ which provide evidence for a sizeable spin-lattice coupling. These findings highlight a strong mutual interaction between the charge, spin and lattice degrees of EuIn$_2$As$_2$ that may also affect the predicted topological states and thus should be considered in the theoretical predictions and interpretation of experimental data.

%
%
\section{Experimental methods}
High-quality single crystals of EuIn$_2$As$_{2}$ with plate-like shapes have been synthesized with a self-flux method~\cite{Zhang2020PRB}. The in-plane resistivity exhibits a metallic temperature dependence, albeit with a cusp-like maximum around $T_N \simeq$ 18 K [see Fig.~\ref{Fig3}(b)]. The in-plane reflectivity $R(\omega)$ spectra were measured at a near-normal angle of incidence using a Bruker VERTEX 70v FTIR spectrometer with an \emph{in situ} gold overfilling technique~\cite{Homes1993}. Data from 30 to 20\,000\icm\ were collected at different temperatures with the sample mounted in an ARS-Helitran crysostat. The room temperature optical response function in the near-infrared to ultraviolet range (4\,000 -- 50\,000\icm) was measured with a commercial ellipsometer (Woollam VASE). The optical conductivity and related response functions and constants were obtained by performing a Kramers-Kronig analysis of $R(\omega)$~\cite{Dressel2002}. For the low frequency extrapolation below 30\icm, we used a Hagen-Rubens function ($R = 1 - A\sqrt{\omega}$). On the high frequency side, the extrapolation was anchored by the room temperature ellipsometry data.

%
%

\section{Results and discussions}
\subsection{Plasma edge and screened plasma frequency}

\begin{figure}[tb]
\includegraphics[width=\columnwidth]{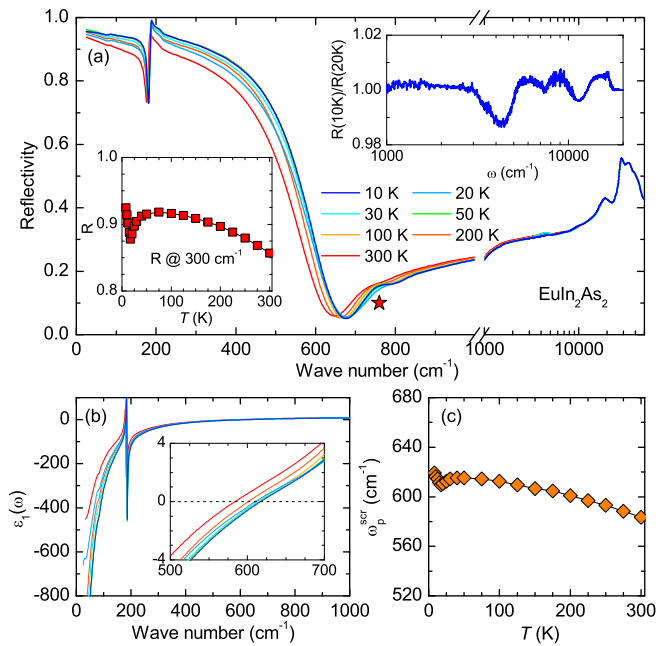}
\caption{ (color online) (a) Temperature dependent in-plane reflectivity spectra of EuIn$_2$As$_2$. The left inset shows the temperature-dependence of the reflectivity at 300\icm. The right inset displays the change of the reflectivity below $T_N \simeq$ 18~K in terms of the ratio of the spectra at 10 and 20~K. (b) Temperature dependence of the real part of the dielectric function $\varepsilon_1(\omega)$. Inset: magnified view of the $\varepsilon_1(\omega)$ spectra in the vicinity of the zero crossing. (c) Temperature evolution of the screened plasma frequency, $\omega^{\rm scr}_{p}$ as deduced from the zero crossing of $\varepsilon_1(\omega)$.}
\label{Fig1}
\end{figure}
Figure~\ref{Fig1}(a) shows the temperature-dependent spectra of the in-plane reflectivity, $R(\omega)$, of EuIn$_2$As$_2$. In the far-infrared range they show a typical metal-like response with a sharp plasma edge, below which the reflectivity increases rapidly and approaches unity toward the origin. The left inset details the temperature dependence of the low-frequency value $R(\omega = 300\icm)$ which exhibits a pronounced anomaly around $T_N \simeq$ 18~K. The small value of the plasma edge ($\sim 650$~\icm) suggests a rather low carrier density, consistent with the small hole pocket that has been observed in ARPES measurements~\cite{Regmi2020PRB,Zhang2020PRB,Sato2020PRR}. The slight temperature dependent upward shift of the plasma edge from around 650\icm\ at 300~K to 680\icm\ at 10~K indicates a corresponding weak increase of the plasma frequency with cooling. Note that the weak feature around 750\icm\ (marked by a star) is due to a plasmonic effect that will be discussed elsewhere. The far-infrared spectra also show a pronounced infrared-active phonon mode around 185\icm\ and there seem to be two additional, very weak modes around 80\icm\ and 215\icm. Toward higher frequency, starting from the mid-infrared range, the spectra reveal several interband transitions from the occupied states in the valence bands to the empty states in the conduction bands, that show up as kinks or peaks that are centered around 5\,000, 12\,000, 20\,000, 27\,500 and 32\,000\icm. The inset on the right-hand side of Fig.~\ref{Fig1}(a) shows the anomalous change of the reflectivity spectra in the AFM state, in terms of the ratio of the spectra at 10 and 20~K. It reveals characteristic peak-dip-peak features around 5\,000 and 12\,000\icm\ that arise from the magnetic splitting of the bulk valence bands and will be further discussed below.

\begin{figure*}[tb]
\includegraphics[width=\textwidth]{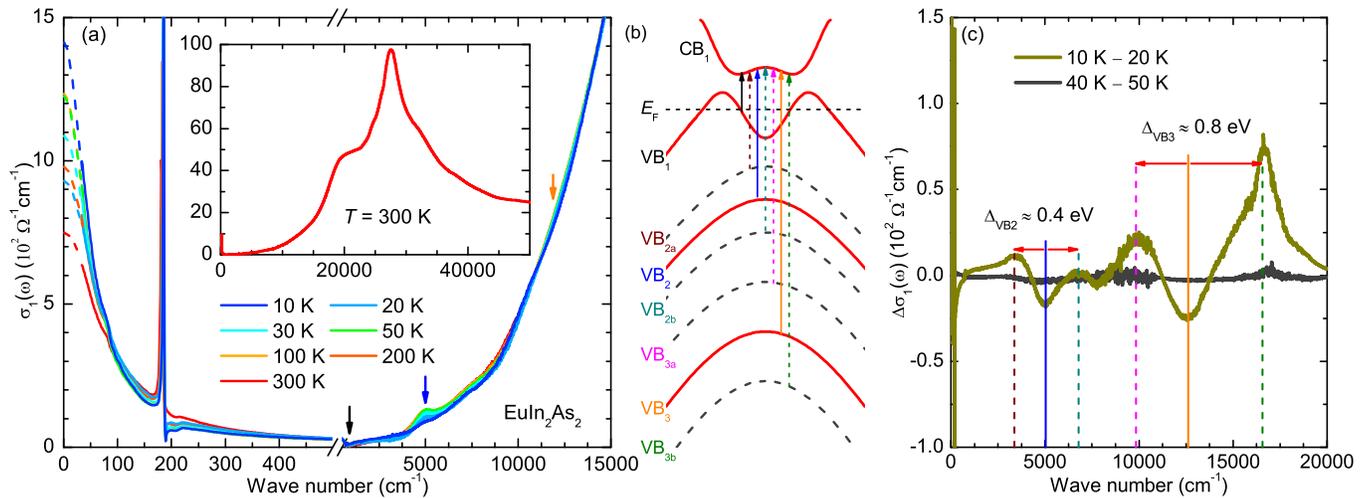}
\caption{ (color online) (a) Optical conductivity of EuIn$_2$As$_{2}$ at different temperatures; the dashed line show the extrapolation of the Drude fits; Inset: Optical conductivity at room temperature in the full measured range up to 50\,000\icm. (b) Schematic of the band structure of EuIn$_2$As$_{2}$. The dashed lines denote the splitting of the valence bands below $T_N$. (c) Difference plot of $\sigma_1(\omega)$ across (10~K $-$ 20~K) and above (40~K $-$ 50~K) the AFM transition.}
\label{Fig2}
\end{figure*}
Figure~\ref{Fig1}(b) shows the corresponding temperature dependent the spectra of the real part of the dielectric function $\varepsilon_1(\omega)$. At low frequencies, $\varepsilon_1(\omega)$ is negative (a defining property of a metal) and can be well described with a Drude model, $\varepsilon(\omega) = \varepsilon_{\infty} - \frac{\omega_p^2}{\omega^2 + i\omega/\tau}$, where $\varepsilon_{\infty}$ is the high-frequency dielectric constant, $\omega_p = \sqrt{ne^2/\epsilon_0 m^\ast}$ is the plasma frequency, that is a measure of the ratio of the carrier density $n$ and the effective mass $m^\ast$ of the free carriers, and $1/\tau$ is their scattering rate. The zero crossing of $\varepsilon_1(\omega)$ (indicated by the horizontal dashed line in the inset) marks the screened plasma frequency $\omega_p^{\rm scr}$ of the free carriers, which is related to the plasma frequency through $\omega_p^{\rm scr} = \omega_p/\sqrt{\varepsilon_{\infty}}$. Fig.~\ref{Fig1}(c) shows that, similar to the low-frequency reflectivity in the left inset of Fig.~\ref{Fig1}(a), $\omega_p^{\rm scr}$ increases weakly toward low temperature, from about 580\icm\ at 300~K to 620\icm\ at 10~K, and exhibits an anomalous suppression around $T_N$.

\subsection{Optical conductivity and band splitting in the AFM state}

Fig.~\ref{Fig2}(a) displays the temperature dependence of the real part of the optical conductivity $\sigma_1(\omega)$ in the far-infrared to near-infrared range. The inset shows the room temperature spectrum of the optical conductivity over the full measured range up to 50\,000\icm. Below 500\icm\, the main features are a weak Drude peak centered at zero frequency and a sharp infrared-active phonon around 185\icm. The onset of the interband transitions occurs around 800\icm\ and is superimposed on the tail of the Drude peak. Towards higher frequency, there is a series of weak interband transitions up to 15\,000\icm\ that is followed by three much stronger interband transitions around 20\,000, 27\,500 and 32\,000\icm.

Fig.~\ref{Fig2}(b) shows a schematic of the band structure that is motivated by the reported band structure calculations~\cite{Xu2019PRL}. The assignment of the low-energy interband transition is indicated with colored arrows. In addition to a pair of conduction and valence bands (CB$_1$ and VB$_1$), for which the spin-orbit-coupling gives rise to an inverted band gap, there are two more valence bands (VB$_2$ and VB$_3$) that are degenerate around the $\Gamma$ point in the paramagnetic state (solid lines) and are expected to exhibit a magnetic splitting in the AFM state (dashed gray lines). This band assignment accounts for the quasi-linear conductivity in the frequency range 800 -- 3\,000\icm\ in terms of the interband transitions near the spin-orbit inverted gap (black arrow). The approximately linear frequency-dependent increase of the conductivity between 800 and 3\,000\icm\ agrees well with the presence of 3D linear bands near the Fermi level, that are also apparent from recent ARPES studies~\cite{Regmi2020PRB,Zhang2020PRB,Sato2020PRR}. The onset frequency of 800\icm\ provides an estimate of the spin-orbit gap of about 100~meV. A somewhat lower gap value will be obtained if the so-called Burstein-Moss shift, due to the Pauli-blocking effect by the holes in the valence band, is included. However, since the Drude peak is very weak, the latter effect should be rather small and thus has not been further considered. The absorption peak around 5\,000\icm\ is assigned to the interband transitions from VB$_2$ to the empty states in CB$_1$, as illustrated by the blue arrow in Fig.~\ref{Fig2}(c). The rapid increase of $\sigma_1(\omega)$ around 12\,000\icm\ can be understood in terms of the transitions from VB$_3$ to CB$_1$, as indicated by the orange arrow. Finally, the much stronger and sharp peaks around 20\,000 and 30\,000\icm\ are assigned to transitions from deeper valence bands that are also predicted by the band calculations~\cite{Xu2019PRL}.

Next we focus on the changes of the optical response across the AFM transition of EuIn$_2$As$_{2}$. Fig.~\ref{Fig2}(c) shows the $\sigma_1(\omega)$ difference spectrum between 10 and 20~K across $T_N \simeq 18$~K. It reveals two sets of peak-dip structures that are centered around the interband transitions at 5\,000 and 12\,000\icm, respectively, and are absent in the corresponding difference spectrum between 40 and 50~K. These characteristic peak-dip structures are interpreted in terms of a magnetic band splitting of VB$_2$ and VB$_3$ which is indicated by the dashed lines in Fig.~\ref{Fig2}(b). For the corresponding interband transitions, this magnetic splitting gives rise to a doublet of sub-bands that are located below and above the frequency of the interband transition in the paramagnetic state, respectively, e.g., VB$_{2a}$ $\rightarrow$ CB$_1$, VB$_{2b}$ $\rightarrow$ CB$_1$, VB$_{3a}$ $\rightarrow$ CB$_1$, and VB$_{3b}$ $\rightarrow$ CB$_1$. As shown in Fig.~\ref{Fig2}(c), from this peak-dip structures in the difference spectrum of $\sigma_1(\omega)$ we can estimate the magnitude of the magnetic band splitting, which amounts to $\sim$ 0.4~eV for VB$_2$ and $\sim$ 0.8~eV for VB$_3$. Note that a recent ARPES study found a similar magnetic splitting of VB$_2$ and a reconstruction of VB$_1$ by the emergence of an ``M''-shaped band due to the bulk-band inversion in the AFM state~\cite{Sato2020PRR}. The latter effect is likely weaker and thus not identified in the difference spectrum of the interband $\sigma_1(\omega)$ across $T_N$. The small anomaly of the free carrier plasma frequency around $T_N$, on the other hand, may be an indication of this band inversion.

\subsection{Drude response and spin fluctuations}
\begin{figure}[tb]
\includegraphics[width=\columnwidth]{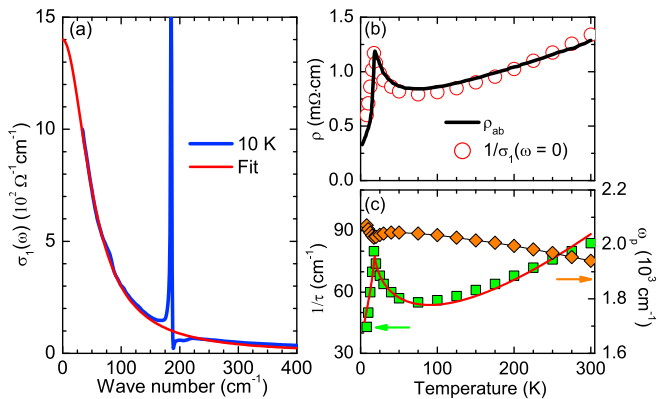}
\caption{ (color online) (a) Drude fit (red line) to the low-frequency optical conductivity of EuIn$_2$As$_{2}$ at 10 K (blue line). Note that the phonon mode at 185\icm\ is not included in this Drude fit. (b) Comparison of the temperature dependence of the dc resistivity, $\rho_{ab}$ (solid line), with the one of the zero-frequency value obtained from the extrapolation of the Drude fit to the optical conductivity, $1/\sigma_1(\omega = 0)$ (open circles). (c) Temperature dependence of the free carrier plasma frequency $\omega_p$ and the scattering rate $1/\tau$, as obtained from the Drude fit.}
\label{Fig3}
\end{figure}
Next we discuss the quantitative analysis of the temperature dependence of the free carrier response that has been obtained from a Drude fit to the low-frequency ${\sigma}_1(\omega)$ spectra. Fig.~\ref{Fig3}(a) shows as an example of the Drude fit to the data at 10~K from which we derive a (bare) plasma frequency of $\omega_p =$ 2056\icm\ and a scattering rate of $1/\tau =$ 50\icm. Corresponding Drude fits of the $\sigma_1(\omega)$ curves have been performed at all measured temperatures, as shown by the dashed lines in Fig.~\ref{Fig2}(a). Fig.~\ref{Fig3}(b) compares the temperature dependence of the dc resistivity $\rho \equiv 1/\sigma_1(\omega = 0)$ from electric transport measurements (black solid line), with the one of the inverse conductivity at zero frequency obtained from the Drude fit (open red circles). They both agree reasonably well concerning the temperature dependence, with a characteristic cusp-like maximum around $T_N$, and even the absolute values. This agreement confirms that the modelling of the optical data is meaningful and reliable. The corresponding plasma frequency $\omega_p$ (orange symbols) and the scattering rate $1/\tau$ (green symbols) are displayed in Fig.~\ref{Fig3}(c). Similar to the screened plasma frequency in Fig.~\ref{Fig1}(c), the bare plasma frequency shows a small increase toward low temperature in the paramagnetic state and only a weak anomaly around $T_N$. The electronic scattering rate, on the other hand, exhibits a pronounced, cusp-like maximum around $T_N$ that is similar to the one of the resistivity $\rho$ in Fig.~\ref{Fig3}(b). In return, this suggests that the cusp-like maximum of the resistivity $\rho = m^{\ast}/e^2 \tau n$ around $T_N$ arises from a corresponding increase of the free carrier scattering rate that is caused by the critical fluctuations of the Eu spins in the vicinity of $T_N$. Our optical data thus provide evidence for a rather strong interaction of the charge carriers with the (slow) spin fluctuations of the Eu moments in EuIn$_2$As$_2$.

To study the nature of the dominant scattering mechanism, we recall the Suezaki-Mori model that was proposed for AFM or order-disorder systems~\cite{Suezaki1968,Suezaki1969,Thomas1973}, where the scattering rate is given by
\begin{equation}
\label{scatteringrate_model}
\frac{1}{\tau(T)}= A + BT +C(1-D|\epsilon|^{2\beta}).
\end{equation}
The first and second terms represent the contributions of impurities or vacancies and of the phonons. The third term $1/\tau_C \propto 1-D|\epsilon|^{2\beta}$ accounts for the critical contribution, which is written in terms of the reduced temperature $\epsilon = (T - T_0)/T_0$ and the critical exponent $2\beta$. Using scaling estimates for an Ising-type model, the exponent should be $2\beta = 0.625$~\cite{KADANOFF1967,Thomas1973}. With this model, we can well reproduce the temperature dependence of scattering rate [red solid line in Fig.~\ref{Fig3}(c)], in particular, the upward-pointing cusp in the scattering rate that is due to the critical fluctuations. The fitted value of the critical exponent of $2\beta = 0.66(4)$ also agree with the prediction. This confirms that the scattering of the free carriers is strongly enhanced by the critical fluctuations of the Eu spins in the vicinity of $T_N$.

\subsection{Spin-lattice coupling}
\begin{figure}[tb]
\includegraphics[width=\columnwidth]{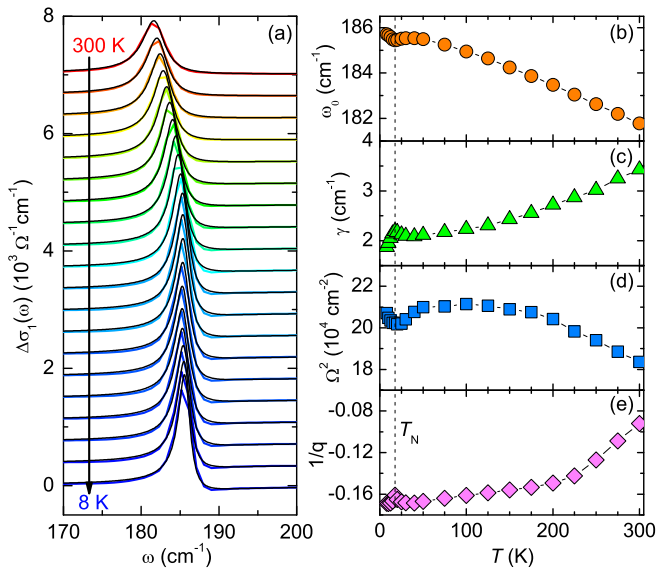}
\caption{ (color online) (a) Line shape of the strong infrared-active phonon mode at temperature from 300 to 8 K in EuIn$_2$As$_2$. The black solid lines through the data denote the Fano fits. (b)--(e) Temperature dependence of the resonance frequency $\omega_0$, the linewidth $\gamma$, the oscillator strength $\Omega^2$, and the Fano parameter $1/q$ of the phonon. The vertical dashed line denotes the antiferromagnetic transition temperature $T_N$ at which all the parameters show anomalies.}
\label{Fig4}
\end{figure}
Finally, we turn to the infrared-active phonon mode at 185\icm. Its temperature dependence is detailed in Fig.~\ref{Fig4}(a) which shows the phonon line shapes after the electronic background (determined with a Drude fit) has been subtracted. With decreasing temperature, this mode shifts to higher frequency and becomes sharper. Meanwhile, the line shape becomes more strongly asymmetric. Such an asymmetric line shape is a characteristic signature of the coupling of the phonon to a much broader background of electronic and/or spin excitations.

The asymmetric phonon mode has been fitted with a Fano-type line shape~\cite{Fano1961} according to the following expression
\begin{equation}
\label{Fano}
\sigma_{1}(\omega)=\frac{2\pi}{Z_0} \frac{\Omega^2}{\gamma}
\frac{q^2 +\frac{4q (\omega - \omega_0)}{\gamma} -1}{q^2 (1 + \frac{4(\omega - \omega_0)^2}{\gamma^2})},
\end{equation}
where $\omega_{0}$, $\gamma$ and $\Omega$ are the resonance frequency, line width, and the strength of the phonon, respectively. The dimensionless parameter $q$ describes the asymmetry of the Fano profile. The parameter $1/q^2$ is a measure of the strength of the coupling between phonon and electron or spin. At $1/q^2 = 0$ the line shape is still symmetric and Lorentzian, whereas with increasing $1/q^2$ the line shape becomes more asymmetric. The best fits using this Fano model describe the phonon mode reasonably well, as shown by the black solid lines in Fig.~\ref{Fig4}(a). The temperature dependence of the fit parameters for $\omega_{0}$, $\gamma$, $\Omega^2$ and $1/q$ are summarized in Figs.~\ref{Fig4}(b) to \ref{Fig4}(e), respectively. They reveal some weak, but clearly resolved anomalies in the vicinity of $T_N \simeq$ 18~K which suggest that the spin-phonon coupling is also not negligible.

%
%
\section{Conclusions}
To summarize, the optical conductivity of the antiferromagnetic axion topological insulator candidate EuIn$_2$As$_2$ has been measured over a wide frequency range and at a variety of temperatures. In the far-infrared range, we observed a weak Drude response and a sharp infrared-active phonon mode. Towards higher frequency, there is a series of interband transitions that starts with a weak absorption edge around 800\icm\ and is followed by weak bands at 5\,000 and 12\,000\icm\ and strong bands at 20\,000, 27\,500 and 32\,000\icm. Based on reported band structure calculations, we assign the weak bands to transitions from the low-lying valence bands (VB$_2$ and VB$_3$) to an empty conduction band (CB$_1$). The gap magnitude is estimated to be 0.1~eV. Below $T_N \simeq 18$~K, we observed clear signs of a magnetic splitting of the valence bands VB$_2$ and VB$_3$ which amounts to about 0.4 and 0.8~eV, respectively. The free carrier response is also strongly affected by the antiferromagnetic transition. In particular, the scattering rate (width of the Drude peak) shows a pronounced, cusp-like maximum at $T_N$ that arises from critical antiferromagnetic fluctuations of the Eu spins that are strongly interacting with the charge carriers. An anomalous $T$-dependence in the vicinity of $T_N \simeq 18$~K is also seen for the infrared-active phonon mode around 185\icm\ which suggests that the spin-phonon coupling is also sizeable. Our study highlights that EuIn$_2$As$_2$ is of interest not only for its magnetic and topological properties, but also for the rather strong interaction amongst its charge, spin and lattice degrees of freedom. The latter may also affect the various topological surface states and need to be considered in comparing theory with experiments.

%
%
\begin{acknowledgments}
Work at the University of Fribourg was supported by the Schweizerische Nationalfonds (SNF) by Grant No. 200020-172611. B.S. acknowledges the support of the Fundamental Research Funds for the Central Universities, Grant No. 19lgpy260.
\end{acknowledgments}

%
%

\end{document}